# Interplay between localization and magnetism in (Ga,Mn)As and (In,Mn)As

Ye Yuan[1,2,*], Chi Xu[1,2], René Hübner[1], Rafal Jakiela[3], Roman Böttger[1], Manfred Helm[1,2], Maciej Sawicki[3], Tomasz Dietl[3,4,5], and Shengqiang Zhou[1]

[1]Helmholtz-Zentrum Dresden-Rossendorf, Institute of Ion Beam Physics and Materials Research, Bautzner Landstrasse 400, D-01328 Dresden, Germany

[2]Technische Universität Dresden, D-01062 Dresden, Germany

[3]Institute of Physics, Polish Academy of Sciences, Aleja Lotnikow 32/46, PL–02668 Warsaw, Poland

[4]International Research Centre MagTop, Aleja Lotnikow 32/46, PL–02668 Warsaw, Poland

[5]WPI-Advanced Institute for Materials Research, Tohoku University, Sendai, 980-8577, Japan

* y.yuan@hzdr.de

# Abstract:

Ion implantation of Mn combined with pulsed laser melting is employed to obtain two representative compounds of dilute ferromagnetic semiconductors (DFSs): $Ga_{1-x}Mn_xAs$ and $In_{1-x}Mn_xAs$. In contrast to films deposited by the widely used molecular beam epitaxy, neither Mn interstitials nor As antisites are present in samples prepared by the method employed here. Under these conditions the influence of localization on the hole-mediated ferromagnetism is examined in two DFSs with a differing strength of *p-d* coupling. On the insulating side of the transition, ferromagnetic signatures persist to higher temperatures in $In_{1-x}Mn_xAs$ compared to $Ga_{1-x}Mn_xAs$ with the same Mn concentration *x*. This substantiates theoretical suggestions that stronger *p-d* coupling results in an enhanced contribution to localization, which reduces hole-mediated ferromagnetism. Furthermore, the findings support strongly the heterogeneous model of electronic states at the localization boundary and point to the crucial role of weakly localized holes in mediating efficient spin-spin interactions even on the insulator side of the metal-insulator transition.

## I. INTRODUCTION

One of the most specific features of magnetic semiconductors is the coexistence of strong exchange coupling effects between carriers and localized spins with intriguing phenomena of quantum localization in disordered systems. There issues are particularly relevant to dilute ferromagnetic semiconductors (DFSs) in which carriers mediate ferromagnetic coupling and, at the same time, are subject to localization [1-14]. The localization in (Ga,Mn)As can be tuned, *e.g.,* by varying Mn concentration [1], by isovalent-anion substitution [15] and by nonmagnetic compensation via codoping [16]. The understanding of the interplay between ferromagnetism and carrier localization remains in a nascent stage and contradicting approaches are under consideration [17-19]. One of the reasons is due to a strong dependence of localization and key magnetic properties on the concentration of poorly controlled donor defects, such as - in the most thoroughly studied system, i.e., (Ga,Mn)As - Mn interstitials [20] and As antisites [21]. Another reason for the slow progress towards a consensus is the intricate nature, even in non-magnetic semiconductors, of the metal-insulator transition (MIT). In particular, characteristic length scales are too large to allow the MIT to be treated by available *ab initio* methods, whereas theoretical tools, such as the renormalization group formalism, provide merely critical exponents and quantum corrections brought about by diffusion poles, rather than the absolute values of experimentally available quantities [22].

In this paper we present results of systematic charge transport and magnetic studies on a series of $Ga_{1-x}Mn_xAs$ films, together with magnetic investigations on $In_{1-x}Mn_xAs$ layers. Both kinds of materials are obtained by Mn ion implantation followed by subsequent pulsed laser melting. Neither Mn interstitials nor As antisites are present in samples prepared in this way [23]. Under these rather unique conditions we explore the interplay between magnetism and quantum localization in the Mn concentration range *x* from 0.3 to 1.8%, which covers both sides of the MIT. We demonstrate in a quantitative fashion how the system evolves with *x* from a paramagnetic (PM) phase (probed down to 1.8 K), to a superparamagnetic (SPM) material, to reach, *via* a mixed phase consisting of percolating ferromagnetic clusters and superparamagnetic grains, a global ferromagnetism (FM) without any superparamagnetism. The absence of superparamagnetism for $x \geq 1.4\%$ makes our samples, grown by PLM, different from those obtained by molecular beam epitaxy (MBE), in which the measurement procedure employed here reveals often a superparamagnetic component even in films with higher *x* [9,12]. Furthermore, the absence of compensation allows us to determine the hole concentration directly from *x*, which provides a solid ground to test the *p-d* Zener model quantitatively. We find excellent agreement between our experimental data and the theoretical prediction [2]. A worthwhile finding in our work is a clear demonstration that the *p-d* interaction enhances the hole localization and, thus, diminishes hole-mediated coupling. This results in a *weaker* ferromagnetic signature in the range of low Mn concentrations in (Ga,Mn)As compared to (In,Mn)As in which hole localization is weaker.

## II. EXPERIMENT

The (Ga,Mn)As and (In,Mn)As samples for this study were prepared by Mn ion implantation into semi-insulating GaAs and intrinsic InAs wafers, respectively, followed by subsequent pulsed laser melting (PLM). The implantation energy was 100 keV, and the wafer normal was tilted by 7° with respect to the ion beam to avoid channeling. According to the stopping and range of ions in matter (SRIM) simulation, the longitudinal straggling ($\Delta R_P$) for the Mn distribution in GaAs and InAs is around 31 and 38 nm, respectively. A Coherent XeCl laser (with 308 nm wavelength and 28 ns pulse duration) was employed to recrystallize the samples, and the energy densities were optimized to achieve both the highest crystalline quality and the best randomization of the Mn distribution: 0.3 J/cm$^2$ for (Ga,Mn)As and 0.2 J/cm$^2$ for (In,Mn)As [24].

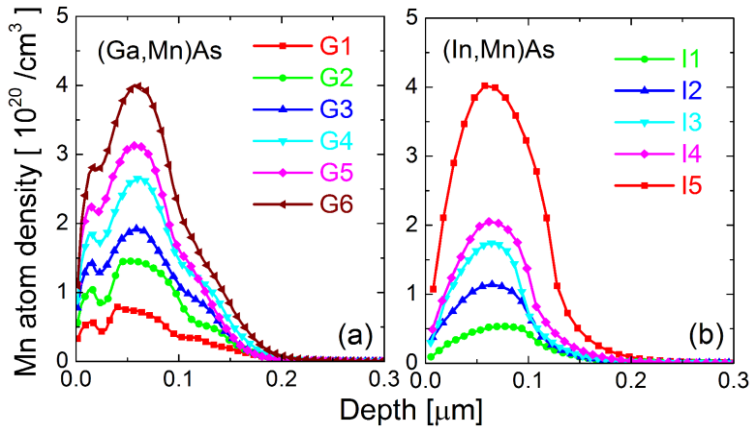

Fig. 1. Concentration profiles of Mn in (Ga,Mn)As and (In,Mn)As determined by SIMS measurements.

Mn concentration profiles were determined by secondary ions mass spectrometry (SIMS) technique using Cameca IMS 6F microanalyser. SIMS measurement was performed with the cesium (Cs$^+$) primary beam. Mn concentrations were derived from the intensity of MnCs$^+$ clusters. Since the Mn distribution in both (Ga,Mn)As and (In,Mn)As is approximately Gaussian the Mn concentration relevant for the measured $T_C$ is taken as an average value within the coherence length (which is of the order of 5 nm) in the region around the maximum, as $T_C$ is determined by the peak Mn concentration in the distribution [9,15].

Magnetic properties were studied by employing a Quantum Design MPMS XL Superconducting Quantum Interference Device (SQUID) magnetometer equipped with a low field option. For the thermo-remnant magnetization (TRM – the temperature dependence of the remnant magnetization measured upon warming) measurements, the samples were cooled down under a field of 1 kOe, then at the base temperature the field was switched off using a soft quench of the superconducting magnet and the system was warmed up while collecting data. When above the magnetic critical temperature ($T_C$ – taken here as the temperature where the TRM vanishes), the samples were re-cooled to the starting temperature at the same zero-field conditions while the data recording was continued and entitled as spontaneous magnetization $M_S$. All the magnetic measurements were carried out using an about ~20 cm long silicon strip to fix the samples and the adequate experimental code for minute signal measurements was strictly observed [25]. Temperature dependent transport measurements were carried out using van der Pauw geometry in a Lakeshore Hall measurement system.

## III. RESULTS AND DISCUSSION

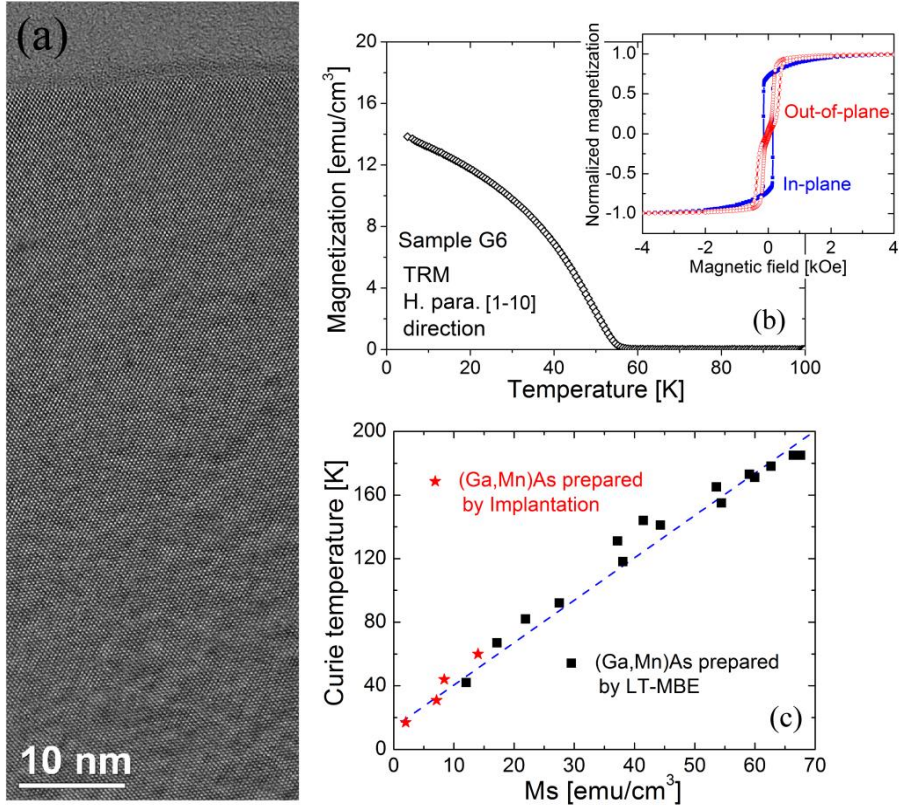

Fig. 2. (Color online) Structural (a) and magnetic properties (b,c) of (Ga,Mn)As epilayers prepared by ion implantation and PLM. (a) A cross-sectional high-resolution TEM image of the (Ga,Mn)As sample with 1.8% Mn points to high crystalline quality and excludes the presence of any extended lattice defects, amorphous inclusions, and precipitates of other crystalline phases. Temperature dependence of magnetization (b), the character of magnetic anisotropy [inset to (b)], and the magnitude of Curie temperature at given spontaneous magnetization $M_S$ in our (Ga,Mn)As follow the trend established in optimized (Ga,Mn)As films grown by LT-MBE [26].

We have prepared both (Ga,Mn)As and (In,Mn)As with very low Mn concentrations, as shown in Tab. I. As shown in Fig. 2a, the perfect lattice-fringe image in the cross-sectional high-resolution transmission electron microscopy (HR-TEM) indicates that PLM leads to the complete epitaxial recrystallization of the implanted region even if $x$ is as high as 1.8%. Importantly, for the same sample, a concave curvature of TRM indicates a nearly mean-field theory behavior, as shown in Fig. 2b. More convincing evidence to support the epitaxial nature of (Ga,Mn)As on the GaAs substrate is the character of magnetic anisotropy, as shown in the inset of Fig. 2b. Due to the compressive strain in the (Ga,Mn)As epilayer, an in-plane magnetic easy axis is observed, as expected on the ground of the Zener model and typically observed in (Ga,Mn)As.

The magnitude of Curie temperature ($T_C$) in (III,Mn)V DSFs is expected to increase with the hole density $p$ and the effective Mn concentration $x_{\text{eff}}$ [2]. The value of $p$ is controlled by concentrations of substitutional Mn acceptors and compensating donors. In spatially uniform systems, $x_{eff}$ is directly determined by the spontaneous magnetization $M_S$, and is typically smaller than $x$ due to antiferromagnetic interactions, for instance, between substitutional and interstitial Mn ions [15,18]. The determined values of $x$ and $T_C$ are summarized in Table I. In order to compare our $T_C$ values to $T_C(M_S)$ obtained for optimized thin MBE (Ga,Mn)As films [26,27] we take $M_S \sim xN_0m_{Mn}$, where $N_0$ is the cation concentration and $m_{Mn}$ = 4.0 $\mu_B$ in the case of weak compensation, small magnitude of the hole orbital moment [28,29], and large spin polarization of the hole liquid. As indicated in Fig. 2c our (Ga,Mn)As samples that show FM characteristics follow the $T_C(M_S)$ trend established for thin films obtained by MBE and low-temperature annealing [26].

TAB. I. The Mn concentration $x$, Curie temperature $T_C$, and characteristic temperatures of ferromagnetic grains $T_\sigma$ of the $Ga_{1-x}Mn_xAs$ samples (denoted by G) and the $In_{1-x}Mn_xAs$ samples (denoted by I).

| Sample No. | Mn concentration (%) | $T_C$ ($T_\sigma$) (K) |
|---|---|---|
| G1 | 0.35 | 0 |
| G2 | 0.66 | (7.5) |
| G3 | 0.87 | 17 (13) |
| G4 | 1.2 | 31 |
| G5 | 1.4 | 44 |
| G6 | 1.8 | 60 |
| I1 | 0.30 | 0 |
| I2 | 0.63 | (6) |
| I3 | 0.96 | 14 (11) |
| I4 | 1.2 | 23 |
| I5 | 2.2 | 40 |

In doped semiconductors the critical carrier concentration corresponding to the MIT is usually well described by the Mott formula [17,22]:

$$p_c^{\frac{1}{3}} a_B = 0.26 \pm 0.05 \qquad (3)$$

$$a_B = \frac{e^2}{8\pi\varepsilon_0\varepsilon_r E_I} \qquad (2)$$

where $a_B$ is the effective Bohr radius, $p_c$ is the critical hole concentration, $e$ is the charge of single electron, $\varepsilon_r$ is the static dielectric constant, $\varepsilon_0$ is the vacuum permittivity, and $E_I$ is the impurity binding energy. For Mn in GaAs, $\varepsilon_r$ = 12.9, $E_I$ = 112.4 meV, thus a critical concentration of $p_c$ ranging from $0.7 \times 10^{20}$ to $2.4 \times 10^{20}$ cm$^{-3}$ is obtained [17,30]. We compare this theoretical value to the hole concentrations in our samples assuming that each substitutional Mn atom delivers one hole, $p \cong xN_0$, where $N_0 = 2.2 \times 10^{22}$ cm$^{-3}$ is the cation density in GaAs. The absence of compensation was proved by Rutherford backscattering channelling which showed that (Ga,Mn)As films prepared by ion implantation and PLM are free from Mn interstitial defects [23]. Moreover, owing to the high temperature nature of PLM, the formation of arsenic antisite defects can also be excluded [26]. Under these rather unique conditions we find that the values of $p$ spans from $8.8 \times 10^{19}$ to $4 \times 10^{20}$ cm$^{-3}$ in samples G1 to G6, respectively. This means, in agreement with our resistance measurements discussed below, that our samples are probing both sides of the MIT.

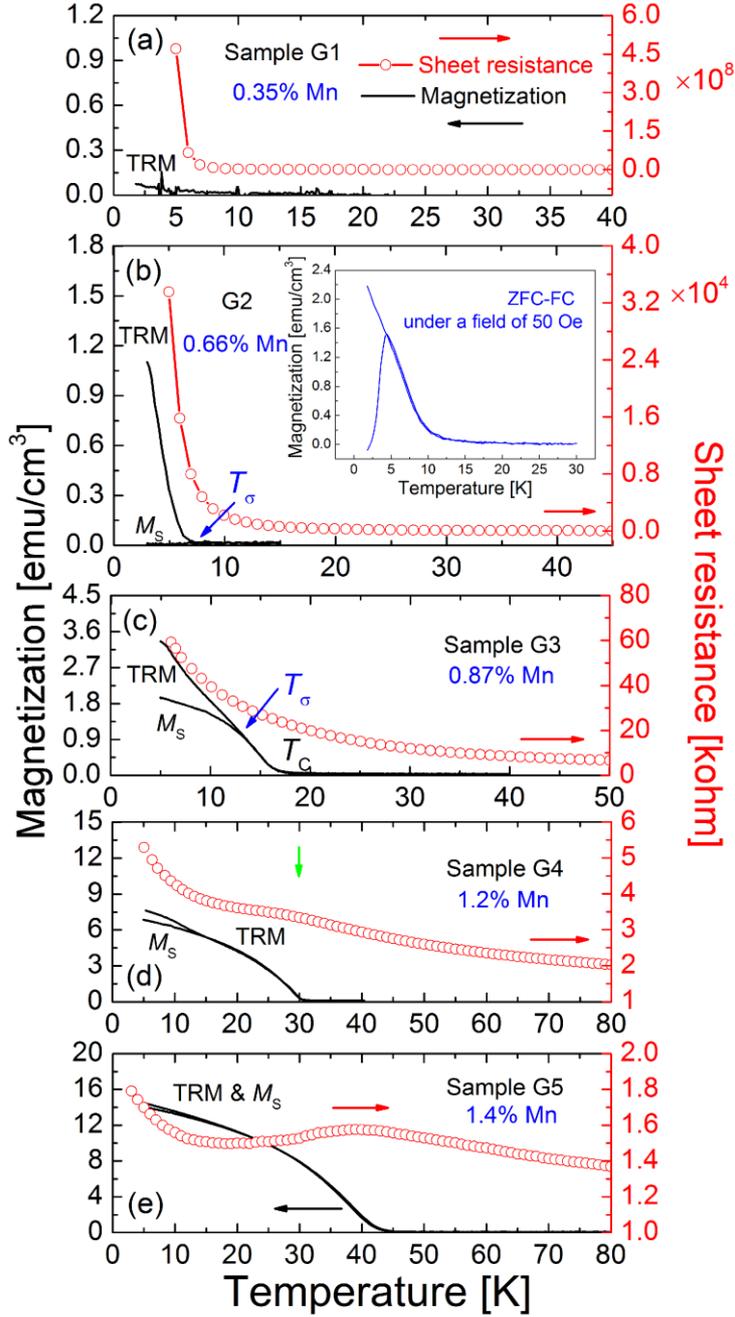

Fig. 3. (Color online) The interplay between localization and magnetism in (Ga,Mn)As in the vicinity of metal-insulator transition. Temperature dependence of remnant magnetization (lines, left axis) and sheet resistance (circles, right axis) of (Ga,Mn)As (samples G1-G5). The inset to (b) shows ZFC and FC curves for sample G2 in a field of 50 Oe. Upon increasing Mn concentration, together with the emergence of metallic conductivity, differences between spontaneous magnetization ($M_S$) and thermo-remnant magnetization (TRM) get reduced. This indicates that the long-range global ferromagnetic order gradually replaces the mesoscopic ferromagnetic order when hole localization diminishes.

In Fig. 3, solid lines and open circles represent the temperature dependent TRM and sheet resistance, respectively. The resistance in the GΩ range of sample G1 (Fig. 3a) indicates a robust localization of carriers. The conductivity can be described as variable range hopping *via* a Coulomb gap with a characteristic energy of 3.4 meV in the whole temperature range, as shown in Fig. 4. There is no detectable remnant magnetization in this sample, indicating that the FM coupling does not develop for such a low $x$ value above 2 K - the sample is in a paramagnetic state.

However, clear indications of FM coupling are seen for the remaining samples with x ≥ 0.66%. Namely, all these samples show the existence of TRM, whose thermal properties change significantly with $x$. In particular, upon increasing $x$ the TRM vanishes at progressively higher temperatures. Furthermore, the TRM curvature changes from a convex for sample G2 with $x$ = 0.66%, through a mixed case for sample G3, to a concave one for larger $x$. Remarkably, hand in hand with these changes a spontaneous magnetization $M_S$ becomes visible when the sample is cooled back at the same zero-field conditions under which the TRM was measured. This ferromagnetic response $M_S$: (i) appears on cooling at exactly the same temperature $T_C$ at which the TRM shows up, and (ii) $M_S$ follows the TRM only when the TRM's curvature is concave, otherwise the $M_S$ trails below the TRM. As detailed below, information encoded in TRM and $M_S(T)$ measurements proves sufficient to assess the magnetic constitution of the studied layers.

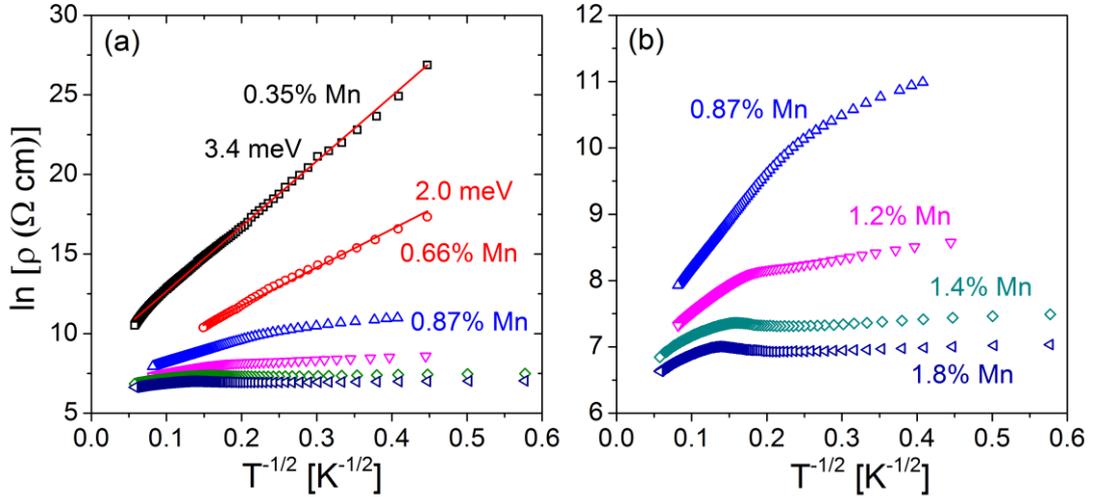

Fig. 4. (Color online) (a) Temperature dependence of resistivity ρ in (Ga,Mn)As in the absence of an external magnetic field. The results are shown as ln(ρ) vs $T^{-1/2}$. In samples G1 and G2, the linear dependence dominates across the whole temperature range, supporting the crucial role of hopping mechanism in the electronic transport. However, in other samples with higher Mn concentrations, the conduction mechanism significantly changes, as magnified in (b).

Sample G2 with $x$ = 0.66% is the lowest-$x$ layer exhibiting a non-zero TRM. This is indicative that FM coupling is present here, but the rapid increase of the resistivity at low temperatures, despite four orders of magnitude lower values than in sample G1, still points to a sizable localization that precludes a long-range (global) ordering mediated by itinerant holes. Indeed, this is the case – the FM coupling is maintained only over a nanometer-range distance.

To substantiate the claim above we resort to low-temperature sample cycling in a weak field of 50 Oe in the well-established protocol of zero-field cooled (ZFC) and the field cooled (FC) manner. The results presented in the inset to Fig. 3b convince us of a granular (nonhomogeneous) magnetic state of this sample, as the magnetic behaviour is typical for blocked SPM ensembles of magnetic particles. In particular, a maximum on the ZFC curve and a clear bifurcation between ZFC and FC data are both seen at nearly the same temperature, corresponding to the (mean) blocking temperature $T_B$ of the ensemble. By using the standard formula for the dynamical blocking, $KV = 25k_BT_B$, where $K$, the anisotropy constant in (Ga,Mn)As, ranges between 5 000 and 50 000 erg/cm$^3$ [30], $V$ is the volume of the magnetic particle, $k_B$ is the Boltzmann constant, and the factor 25 is set by the experimental time scale – about 100 s, in the SQUID magnetometry. This condition implies that $T_B \cong 5$ K corresponds to a sphere of a diameter between 8 to 20 nm, which indeed confirms a mesoscopic extent of FM coupling in this case. Importantly, the appearance of the granular magnetism does not result from nanometer sized Mn aggregates or other types of short scale Mn inhomogeneities, as their presence is excluded by the TEM analyses (see e.g. Fig. 2a). The presence of magnetic particles is assigned to the fact that according to the Anderson-Mott character of the MIT – occurring primarily due to localization of band carriers by scattering – the carriers' localization radius increases only gradually from the Bohr radius in the strong localization limit, $p \rightarrow 0$, towards infinity at the MIT, $p \rightarrow p_c$ [2,6,17,22]. Thus, a magnetic nanoscale phase separation, driven by carrier density fluctuations, is present in the vicinity of the localization boundary. In such a case FM grains are embedded in the PM host background, as observed experimentally [9,12]. At the same time, the presence of randomly oriented nano-sized magnetic grains gives rise to efficient spin-disorder scattering of carriers. This enhances localization at $B = 0$ and leads to a colossal negative magnetoresistance when a magnetic field is applied to polarize the nano-sized ferromagnetic component. Such a colossal negative magnetoresistance has been observed for sample G2 as shown in Fig. 5a and also in donor-compensated (Ga,Mn)As MBE films with higher Mn concentrations [1,32].

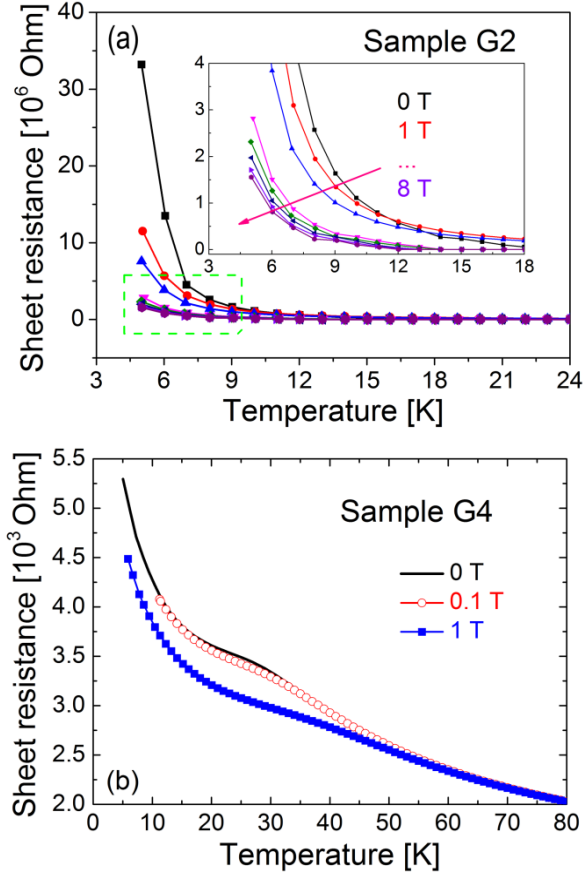

Fig. 5. (Color online) Temperature- and field- dependent resistance in two (Ga,Mn)As samples. In the superparamagnetic sample G2, the resistance at 5 K gets reduced by 95% in 8 T, and the effect results from the suppression of hole scattering by randomly orientated FM grains and orbital quantum localization effect. In the ferromagnetic sample G4, the critical spin-disorder scattering of itinerant holes by fluctuating Mn spins near $T_C$ disappears when the magnetic field is increased to 1T.

Due to the absence of a long-range magnetic coupling in sample G2, the concept of a Curie temperature as the temperature of the thermodynamic phase transition is not appropriate. However, from our measurements we can assess a temperature up to which the magnetic particles survive, $T_\sigma$. From Fig. 3b we get $T_\sigma \cong 7$ K, that is where the TRM vanishes. Finally, we want to point out that no spontaneous moment is observed on cooling at H = 0 across $T_\sigma$. This is yet another strong indication of the mesoscopic scale of the magnetism in this case. On cooling without an external field the magnetic moments of the grains get blocked in random orientations yielding zero net magnetization, although at the remanence it is considerably larger.

Upon increasing the Mn concentration to 0.87%, the global ferromagnetism with a transition temperature $T_C = 17$ K appears, as indicated in Fig. 3c. However, a clear gap which opens between the TRM and $M_S(T)$ below $T_C$, accompanied by a change of the TRM's curvature to a convex one, informs us about the presence of an additional magnetic component possessing similar, superparamagnetic, properties to that observed in sample G2. The value of the TRM - $M_S$ bifurcation temperature, when accompanied by a change to the convex curvature of the TRM, is another practical assessment of $T_\sigma$.

In contrast, global ferromagnetism without any superparamagnetism is found in sample G4 with 1.2% Mn. This is proven by the overlap of curves corresponding to heating and cooling TRM measurements. In this sample a characteristic hump [1, 33] appears in the temperature dependent resistance near $T_C$ = 30 K. In DFSs, the hump comes from critical spin-disorder scattering of itinerant holes by fluctuating Mn spins near $T_C$. Such scattering can be suppressed by an external magnetic field (see Fig. 5b). This sample exhibits a clear increase of the resistivity upon lowering temperature, which points to its insulating character. It means that global FM signatures set in at lower $x$ values, thus at lower hole concentrations than metallic behaviour.

Both metallic behaviour and global ferromagnetism are observed in sample G5 with $x$ = 1.4%. A weak resistance increase at low temperature is related to quantum corrections to conductance on the metallic side of the MIT, associated with disorder-modified carrier-carrier interactions [6,34]. On the other hand, as shown in Fig. 6, the negative magnetoresistance at low temperatures $T << T_C$ in this sample can be well fitted within the single-electron quantum localization scenario [35].

$$\frac{\Delta\rho(B)}{\rho_0} \approx -\frac{\Delta\sigma}{\sigma} = -\frac{n_V e^2 C_0 \rho_0 (eB/\hbar)^{\frac{1}{2}}}{2\pi^2 \hbar} \tag{3}$$

where $C_0 \approx 0.605$, $\rho$ is the resistivity and $\sigma$ is the electrical conductivity, and $n_V/2$ is the number of spin subbands contributing to charge transport. The fitted value $n_V$ = 1.6 indicates that at least three subbands are occupied.

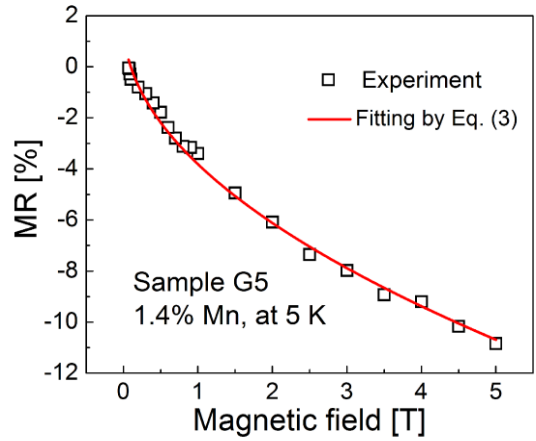

Fig. 6. Quantum localization-induced negative magnetoresistance in metallic (Ga,Mn)As. Negative magnetoresistance is observed at 5 K in sample G5 in the fields in which Mn spins are saturated (open squares). A remarkably good fitting (solid line) suggests that single-carrier orbital weak-localization magnetoresistance dominates at low temperatures.

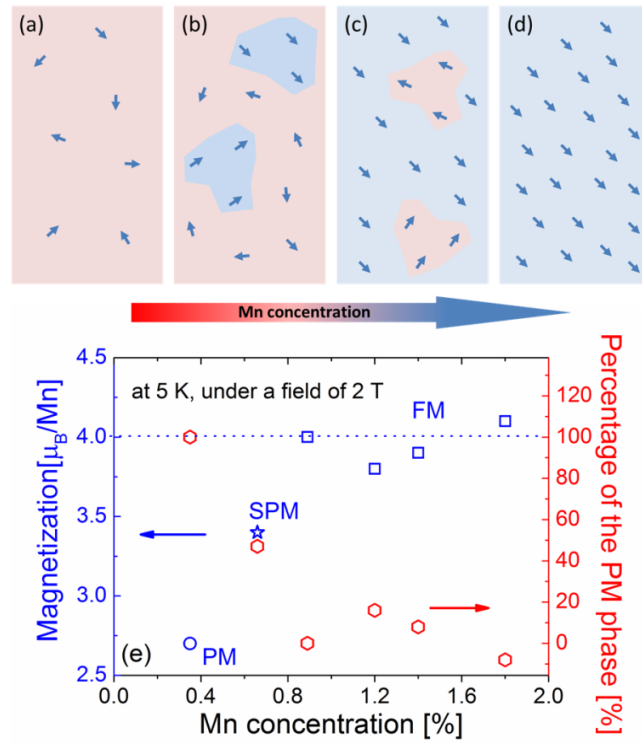

Fig. 7. (Color online) Descriptions of the transition from the paramagnetic to ferromagnetic phases in (Ga,Mn)As. (a-d) A schematic diagram of the evolution of magnetic order in (Ga,Mn)As with increasing Mn concentration. (e) Mn-concentration dependent magnetization in the paramagnetic (circle), superparamagnetic (star), and ferromagnetic (squares) samples measured under a field of 2 T at 5 K. The normalized magnetization per Mn atom was determined by dividing the magnetization by the total number of Mn ions obtained by integrating the distribution of the Mn concentration.

The evolution of magnetism with *x*, as determined for our samples, is illustrated schematically in Figs. 7a-d. The normalized magnetization per Mn atom is calculated from the magnetization divided by the integrated number of all Mn atoms in the layer. The magnetization has been measured at 5 K and 2 T to saturate both the SPM and FM components. The results are displayed in Fig. 7e. They allow us to obtain information on the degree of hole localization and on the relative participation of the ferromagnetic component.

Because of the low Mn and hole concentration $p = 8.8 \times 10^{19}$ cm$^{-3}$ only paramagnetism is observed in sample G1 for which the normalized magnetization is determined to be $M = 2.8$ $\mu_B$/Mn. This value allows us to find out whether the holes are in the strongly or weakly localized regime in this sample. In the former case, the holes are localized on parent Mn acceptors and the magnetic moment per Mn atom can be calculated from the Brillouin function,

$$M = Jg\mu_B[\frac{2J+1}{2J}\coth(\frac{2J+1}{2J}x) - \frac{1}{2J}\coth(\frac{1}{2J}x)] \qquad (4)$$

$$x = \frac{Jg\mu_B H}{k_B T} \qquad (5)$$

where $J = 1$ and $g = 2.77$ [36]. This formula leads to $M = 1.6$ $\mu_B$/Mn at 5 K in 2 T, which implies that the model of strong localization is not applicable for this Mn concentration in question. On the other hand, according to Eq. 4 the $M = 2.7$ $\mu_B$ for $J = 5/2$ and g = 2.0 which is close to the established value of 2.8 $\mu_B$/Mn in sample G1. This is consistent with the fact that in the weakly localized regime, in which holes reside in the valence band, the degree of the hole spin polarization is small, as the Fermi level is about 80 meV below the valence band top whereas the valence band spin splitting is below 12 meV [37].

In the SPM sample G2 $M = 3.4$ $\mu_B$/Mn is obtained, indicating that not all Mn spins contribute to the detectable magnetic moment as depicted in Fig. 7b. Such a value can be used for quantitatively evaluating the electronic phase separation between the ferromagnetic and paramagnetic phases in this (Ga,Mn)As system. For sample G2, $M = 3.4$ $\mu_B$/Mn implies a mixture of the nano-sized hole-rich ferromagnetic phase and the paramagnetic matrix with only very few holes. Therefore, the values of 4 and 2.8 $\mu_B$/Mn in the ferromagnetic and paramagnetic regions, respectively, can be used to quantitatively calculate the composition of each phase according to $4\mu_B(1-y) + 2.8\mu_B y = M_{Mn}$, where *y* is the percentage of the paramagnetic phase, i.e., of the component without ferromagnetic coupling, and *M* is the measured value in Fig. 7e. As the Mn density increases, the percentage of paramagnetic Mn is gradually decreasing: From 100% in sample G1, through 47% in sample G2, finally gets saturated at around 0% in samples G3-G6. The results correspond to the electronic picture given in Figs. 3 and 7, i.e. the inhomogeneity of the ferromagnetism in the sample in the MIT regime directly comes from the electronic phase separation. Note that, since the implanted Mn atoms are not distributed in a rectangular fashion, but exhibit a Gaussian shape, a tail with lower Mn concentrations always exists in all samples. However, only a small part of the superparamagnetic phase is seen through TRM measurements in the ferromagnetic samples G4 and G5 (see in Figs. 3d and e), indicating the tail with low Mn concentration is negligible.

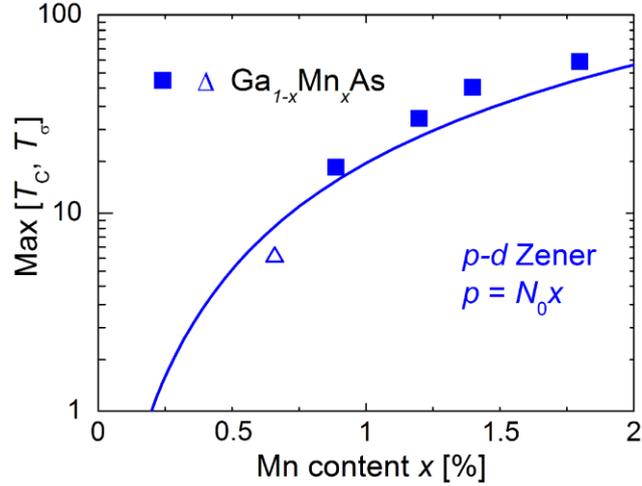

Fig. 8. Curie temperatures $T_C$ (full points) and SP temperatures $T_\sigma$ (empty triangle) in $Ga_{1-x}Mn_xAs$. The solid line shows the prediction by the *p-d* Zener model for $Ga_{1-x}Mn_xAs$ assuming the absence of compensating donors. The dependence of the $T_C(T_\sigma)$ on Mn content matches the Zener model prediction for both SPM and FM (Ga,Mn)As samples.

It is interesting to compare the experimental values of $T_C$ and $T_\sigma$ to the expectations of the *p-d* Zener model. As shown in Fig. 8, there is good agreement between the measured and computed values for both insulating and metallic samples. This finding substantiates the applicability of the *p-d* Zener model for the description of ferromagnetism mediated by itinerant holes as well as by weakly localized holes.

The versatility of ion implantation allows us to compare (Ga,Mn)As with (In,Mn)As. Previous studies of (In,Mn)As films obtained by MBE [38,39] and ion implantation [40,41] with relatively large Mn concentrations show lower $T_C$ compared to (Ga,Mn)As, in agreement with theoretical expectations [2,37]. According to our results presented in Fig. 9, an evolution from the paramagnetic phase to the global ferromagnetic state which takes place in (Ga,Mn)As with increasing *x* is also observed in (In,Mn)As samples obtained by ion implantation and PLM.

Only paramagnetic behaviour is observed in sample I1 with *x* = 0.30% like in the (Ga,Mn)As sample with *x* = 0.35%. In sample I2, both TRM and a bifurcation between ZFC and FC co-imply the SPM character, as in the case of sample G2. When the Mn concentration reaches 0.96%, the TRM measurement points to the coexistence of superparamagnetism and long-range ferromagnetism, which is similar to the case of the (Ga,Mn)As sample with *x* = 0.87%. Upon further increase of *x* to 1.2%, global ferromagnetism dominates, similarly to (Ga,Mn)As with the same *x*. Due to the narrow bandgap nature intrinsic InAs substrates needed for these studies are highly conductive, thus preventing magneto-transport measurements for thin (In,Mn)As layers. However, as established previously [38,39], electrical properties of ferromagnetic (In,Mn)As prepared on insulating GaAs are similar to those of (Ga,Mn)As. The negative magneto-resistance and anomalous Hall effect of (In,Mn)As are observed as in (Ga,Mn)As [38,39,42]. In the Ref. 42, the authors also discussed the possible magnetic phase separation due to inhomogeneous distribution of acceptor impurities.

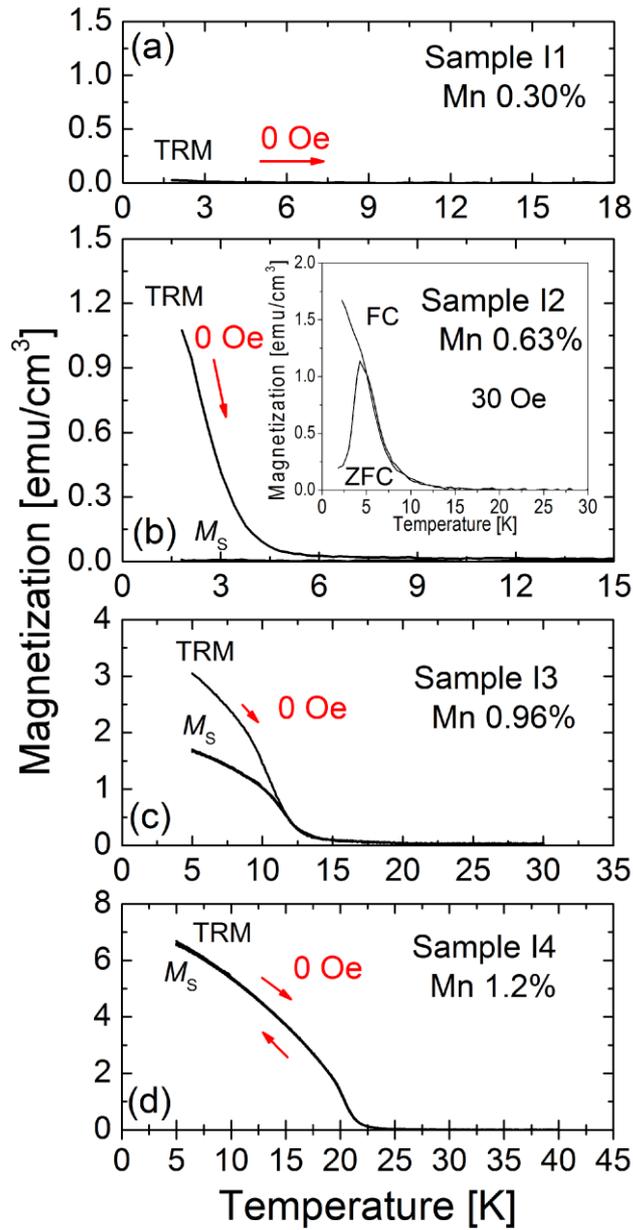

Fig. 9. The transition from the PM, via SPM, to FM phase in (In,Mn)As samples at the MIT regime. Temperature dependent thermo-remnant magnetization of (In,Mn)As samples of (a) I1, (b) I2, (c) I3, and (d) I4. The inset to (b) shows the temperature dependent magnetization under a field of 30 Oe after field cooling (FC) and zero field cooling (ZFC) for sample I2.

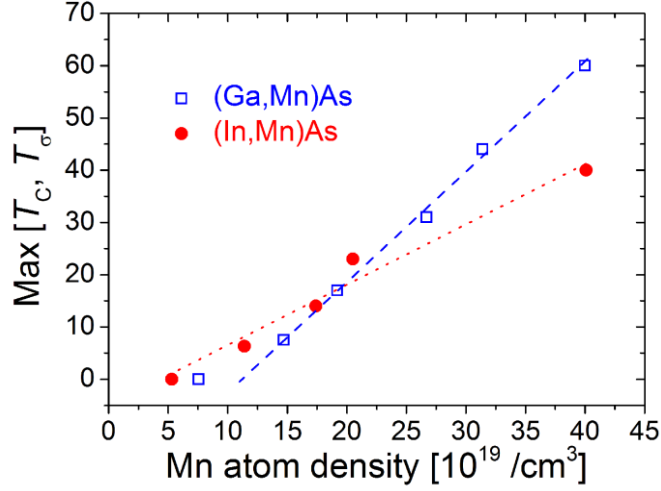

Fig. 10. (Color online) Influence of the *p-d* coupling strength on magnetic properties of (Ga,Mn)As and (In,Mn)As. The dependence of the Curie and superparamagnetic temperature ($T_C$ and $T_\sigma$, respectively) on the Mn atom density for (Ga,Mn)As and (In,Mn)As. The dashed and dotted lines are linear guides for eye for (Ga,Mn)As and (In,Mn)As, respectively. The crossing of two lines, and the higher and lower $T_C/T_\sigma$ in (In,Mn)As at low and high *x* regimes, respectively, result from weaker *p-d* coupling in (In,Mn)As compared to (Ga,Mn)As.

The Curie and SPM temperatures, $T_C$ and $T_\sigma$, respectively of a series of (Ga,Mn)As and (In,Mn)As samples with sequentially increasing Mn densities are shown in Fig. 10. Interestingly, an approximately linear dependence of Max [$T_C$, $T_\sigma$] *vs.* the Mn concentration is found for both materials, however, with differing slopes. In the regime of $N \leq 1.9 \times 10^{20}$ cm$^{-3}$, the (Ga,Mn)As samples exhibit a lower $T_C$ or $T_\sigma$ at given Mn concentrations, i.e., weaker ferromagnetism compared to (In,Mn)As, while in the regime of $N > 1.9 \times 10^{20}$ cm$^{-3}$, higher Curie temperatures are observed in (Ga,Mn)As. This remarkable observation substantiates experimentally the dual role of *p-d* exchange coupling in DFSs, as discussed theoretically previously [43]. Deeper in the metallic regime, a larger *p-d* interaction makes the hole-mediated ferromagnetism stronger, so that $T_C$ in (Ga,Mn)As is higher than in (In,Mn)As, as observed previously [1,26,38,39]. However, in addition to controlling ferromagnetic coupling, a larger *p-d* hybridization shifts the MIT to higher hole concentrations, the effect being stronger in (Ga,Mn)As than in (In,Mn)As in which the bond length is longer. The enhanced hole localization makes ferromagnetic features weaker in (Ga,Mn)As compared to (In,Mn)As in the limit of low hole densities.

The interplay between localization and magnetism of (Ga,Mn)As with high Mn concentrations was also investigated by co-doping either with donors [16] or with isovalent anions [15]. The reduction of $T_C$ was observed together with stronger carrier localization. In our current work, we focus on (Ga,Mn)As and (In,Mn)As with very low Mn concentrations. In addition to the decreased $T_C$ upon enhancing carrier localization [15, 16], we find that the superparamagnetic phase in insulating (Ga,Mn)As and (In,Mn)As is not associated with the presence of compensating donor defects but is an intrinsic property originating, presumably, from the electronic phase separation specific to the Anderson-Mott localization.

## IV. CONCLUSIONS

Through combining systematic studies of electrical and magnetic properties, we have presented experimental evidence supporting the heterogeneous model of electronic states at the localization boundary in (Ga,Mn)As and (In,Mn)As without compensating donors. A transition from an insulating (hopping) to a metallic-like conductance is observed, which is accompanied by a gradual build-up of long-range magnetic coupling, as well as by an increase of the Curie temperature. The *p-d* Zener model prediction is consistent with the measured $T_C$ values in metallic samples as well as with the magnitudes of $T_C$ and $T_\sigma$ on the insulator side of the transition, where the ferromagnetic coupling is mediated by weakly localized holes. Furthermore, in the limit of low Mn concentration the interplay between localization and magnetism results in more robust ferromagnetic signatures in (In,Mn)As compared to (Ga,Mn)As in which the stronger *p-d* coupling enhances localization.

## ACKNOWLEDGMENTS

Support by the Ion Beam Center (IBC) at HZDR is gratefully acknowledged. This work is funded by the Helmholtz-Gemeinschaft Deutscher Forschungszentren (HGF-VH-NG-713). The author Y. Y. thanks financial support by Chinese Scholarship Council (File No. 201306120027). The work in Poland is supported by the Narodowe Centrum Nauki through projects MAESTRO (2011/02/A/ST3/00125) and by the Foundation for Polish Science through the IRA Programme financed by EU within SG OP Programme. Financial support by the EU 7th Framework Programme under the project REGPOT-CT-2013-316014 (EAgLE) and the international project co-financed by Polish Ministry of Science and Higher Education, Grant Agreement 2819/7.PR/2013/2, is also gratefully acknowledged.